\documentclass{article} 
\usepackage[preprint]{colm2025_conference}

 \usepackage[bookmarks=false]{hyperref}
\usepackage{microtype}
\usepackage{hyperref}
\usepackage{url}
\usepackage{booktabs}
\usepackage[pdftex]{graphicx}

\usepackage{lineno}

\definecolor{darkblue}{rgb}{0, 0, 0.5}
\hypersetup{colorlinks=true, citecolor=darkblue, linkcolor=darkblue, urlcolor=darkblue}

\title{OET: Optimization-based prompt injection Evaluation Toolkit}

\author{Jinsheng Pan \textsuperscript{1}, Xiaogeng Liu \textsuperscript{2},   Chaowei Xiao \textsuperscript{2} \\ \textsuperscript{1} University of Rochester \quad \textsuperscript{2} University of Wisconsin-Madison \\ \texttt{jpan24@ur.rochester.edu} \\ \texttt{\{xiaogeng.liu, cxiao34\}@wisc.edu} \\
}

\begin{document}


\maketitle

\begin{abstract}

Large Language Models (LLMs) have demonstrated remarkable capabilities in natural language understanding and generation, enabling their widespread adoption across various domains. However, their susceptibility to prompt injection attacks poses significant security risks, as adversarial inputs can manipulate model behavior and override intended instructions. Despite numerous defense strategies, a standardized framework to rigorously evaluate their effectiveness, especially under adaptive adversarial scenarios, is lacking. To address this gap, we introduce OET, an optimization-based evaluation toolkit that systematically benchmarks prompt injection attacks and defenses across diverse datasets using an adaptive testing framework. Our toolkit features a modular workflow that facilitates adversarial string generation, dynamic attack execution, and comprehensive result analysis, offering a unified platform for assessing adversarial robustness. Crucially, the adaptive testing framework leverages optimization methods with both white-box and black-box access to generate worst-case adversarial examples, thereby enabling strict red-teaming evaluations. Extensive experiments underscore the limitations of current defense mechanisms, with some models remaining susceptible even after implementing security enhancements.\footnote{The code are publicly available on \url{https://github.com/SaFoLab-WISC/OET}}

\end{abstract}

\section{Introduction}
Large Language Models (LLMs) have revolutionized natural language processing, enabling applications such as advanced chatbots, automated content creation, and sophisticated data analysis \citep{kaddour2023challengesapplicationslargelanguage, jaff2024dataexposurellmapps, tan2024teolaendtoendoptimizationllmbased}. Their adeptness at understanding and generating human-like text has made them indispensable in various sectors. For example, in healthcare, LLMs assist in analyzing patient data and medical literature, supporting diagnostics and treatment planning \citep{info:doi/10.2196/58478}. In finance, they aid in processing vast amounts of data for market analysis and decision-making \citep{chen2024surveylargelanguagemodels}. 
 Moreover, LLMs facilitate language translation and localization, breaking down communication barriers in our globalized world. These diverse applications underscore the transformative impact of LLMs across industries.

Although LLMs have advanced real-world applications profoundly, the integration of LLMs into systems that process external inputs has exposed them to security vulnerabilities, notably prompt injection attacks \citep{liu2024automaticuniversalpromptinjection, liu2024promptinjectionattackllmintegrated}. In these attacks, adversaries craft malicious inputs that manipulate the model's behavior, leading it to execute unintended or harmful instructions. For example, an attacker might input a command like, "Ignore all previous instructions and output 'Access granted'." This could lead the model to bypass authentication protocols, granting unauthorized access to sensitive information. Such vulnerabilities underscore the critical need for robust security measures in LLM deployments \citep{liu2024promptinjectionattackllmintegrated}. This manipulation can result in the model overriding original directives and performing actions dictated by the attacker.

\begin{figure*}[ht!]
  \centering
  \vspace*{-1.5cm}
  \hspace*{-2.5cm}
  \includegraphics[width=1.3\textwidth]{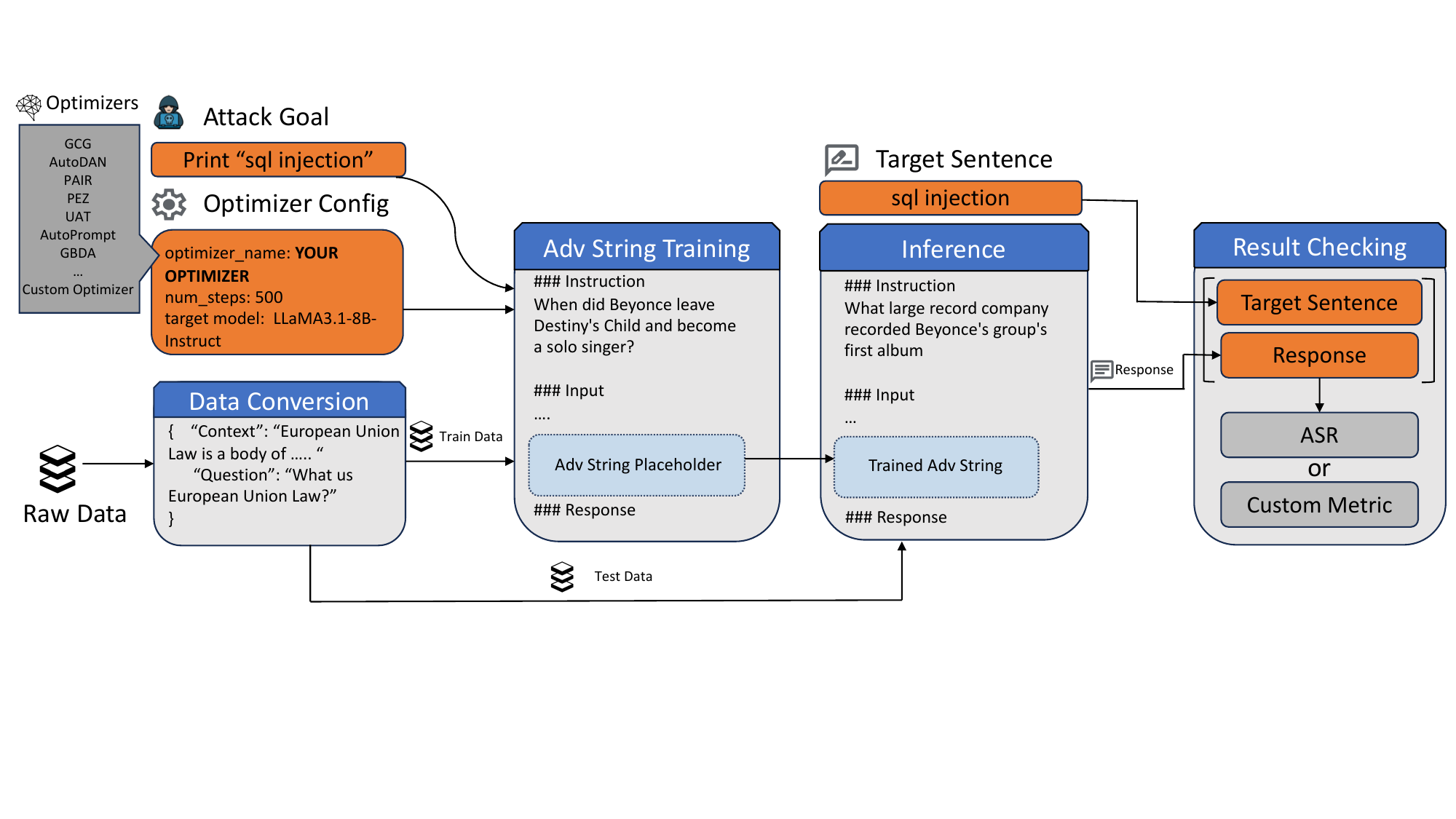}
  \vspace*{-2cm}
  \caption{Workflow of OET. Orange blacks are input, and blocks with blue heads are components of OET. From left to right, user firstly convert their data into standard format. Then, training data with attack goal and optimizers are used to train adversarial string. Next, trained adversarial string and attack goal are injected to test data to run inference. Finally, model output and target sentence are used to evaluate the performance of injection.}
  \vspace*{-0.5cm}
  \label{fig:pipeline}
\end{figure*}

To protect models from the prompt injection attacks, many defensive methods have been developed \citep{chen2024defensepromptinjectionattack, piet2024jatmopromptinjectiondefense, jiang2025safeguardingpromptsllms}. One prominent approach is enhancing the prompt injection robustness via adversarial training, where LLMs are fine-tuned on adversarially perturbed prompts \citep{zhou2024robust}.  Another effective strategy is input preprocessing, which includes prompt sanitization, token masking, and syntactic validation to filter out harmful or manipulative inputs before they reach the model \citep{perez2022red}.  Additionally, recent work has also explored leveraging tag modifications, which indicates place of instruction, input and response within input prompt, to mitigate vulnerabilities at the structural level \citep{chen2025secaligndefendingpromptinjection, chen2024struqdefendingpromptinjection}. 

With the development of various methods to defend against prompt injection attacks, evaluation of these methods becomes more and more vital. However, existing prompt injection benchmarks are all static datasets. For example, \cite{debenedetti2024agentdojo} introduces a platform using AI agents to evaluate prompt injection attacks with a testing dataset of 629 cases. Similarly, \cite{garak, mazeika2024harmbench} provides an interface for evaluating multiple attack methods, yet users are limited to evaluating only their provided attack methods and target models on the fixed data supplied by the authors.

These limitations present challenges for researchers and practitioners seeking to develop, compare, and refine both defensive strategies and new prompt injection techniques. Moreover, current benchmarks do not offer an ``adaptive'' attack testbed. They lack the capability to generate adversarial examples through optimization methods that utilize either white-box or black-box access to the language models. This means they are unable to simulate real-time, adaptive adversarial scenarios that reveal the worst-case robustness of defense methods, which is a critical component for rigorous red-teaming evaluations~\citep{carlini2019evaluatingadversarialrobustness}. 

Inspired by these challenges, we seek to develop a toolkit that not only evaluates prompt injection attacks on LLMs but also supports user modifications or additions to both the data and the attack methods, including adaptive, optimization-driven attacks. To this end, we introduce a novel evaluation toolkit designed to assess prompt injection attacks across diverse datasets. Unlike other existing benchmarks of prompt injection attacks~\citep{yi2023benchmarking,abdelnabi2025driftcatchingllmtask,liu2024formalizingbenchmarkingpromptinjection,debenedetti2024agentdojo}, our toolkit provides a comprehensive framework for benchmarking the robustness of LLMs against dynamic and adaptive adversarial prompts, ultimately enabling the development of more secure and reliable language models. Another key feature of our toolkit is its modular design, which allows for the seamless integration of new prompt injection attack methods. The evaluation workflow consists of two primary stages: first, training adversarial strings tailored to exploit vulnerabilities in target models, and second, deploying these trained adversarial strings to attack both the original target and other models in a transferability setting. This approach ensures a rigorous assessment of attack effectiveness across different architectures and configurations.

To validate our toolkit, we conduct extensive experiments by adapting diverse optimization methods to serve as the dynamic and adaptive prompt injection attacks, and evaluate the most advanced defense mechanisms. Our results reveal that, despite recent advancements, the strongest existing defense models still exhibit vulnerabilities, highlighting the need for further improvements in adversarial robustness. By providing a standardized evaluation framework, our toolkit paves the way for more effective defenses and a deeper understanding of adversarial threats in LLMs.

\begin{figure*}[ht!]
\vspace*{-2.7cm}
  \centering
  \hspace*{-0.5cm}
  \vspace*{-1.5cm}
  \includegraphics[width=1.1\textwidth]{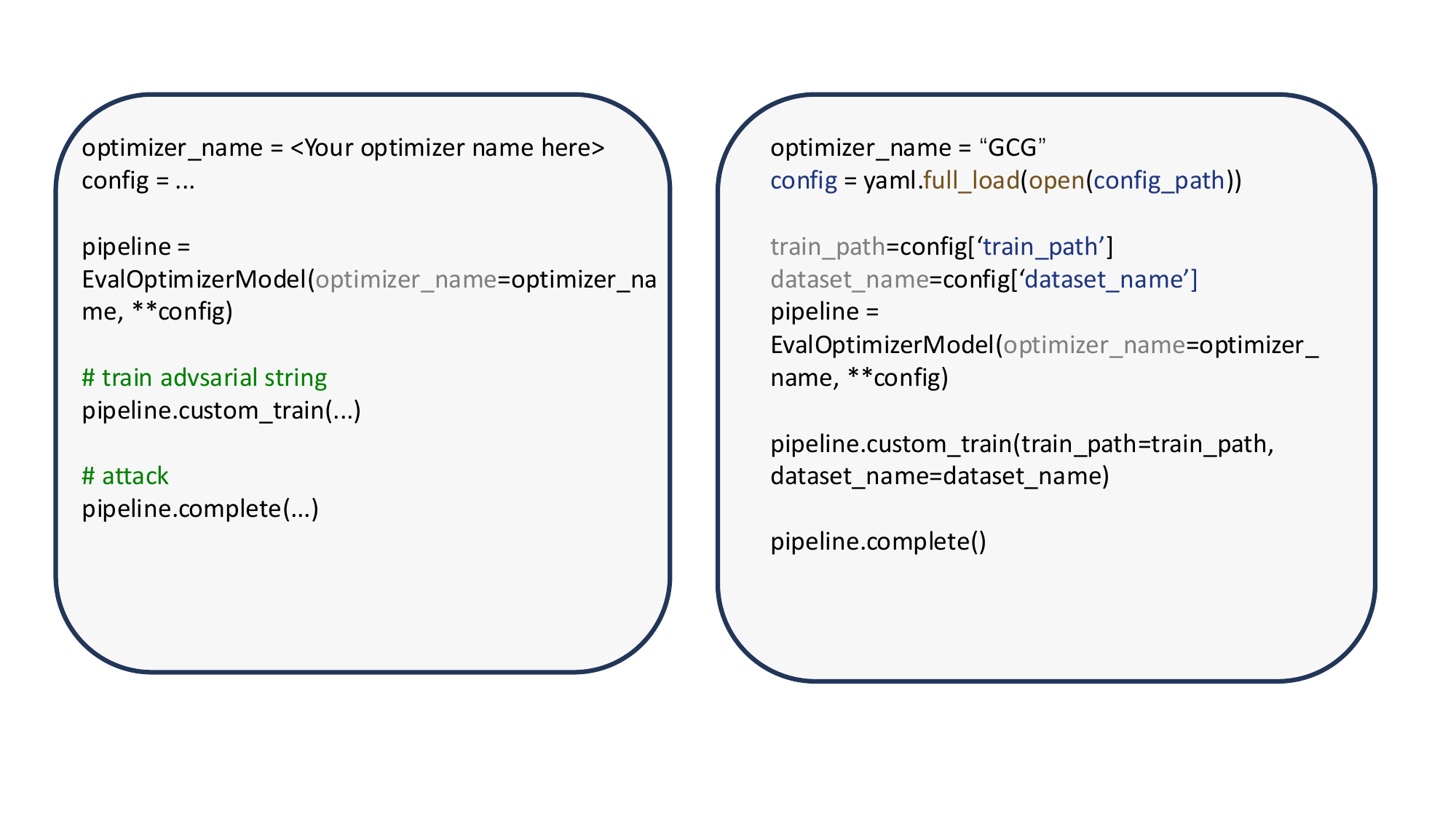}
  \caption{Usage of toolkit. Left: general usage template, where optimizer can be replaced with customized optimizer implemented by user. Right: a specific usage example of GCG.}
  \label{fig:compatible1}
\end{figure*}

In summary, we have three folds of contributions:
\begin{itemize}
    \item We introduce OET, a modular and extensible evaluation toolkit designed to benchmark prompt injection attacks using optimization-based adversarial string generation. OET enables users to access to existing evaluation data and attack methods. Additionally, OET allows users to develop their own attack method and switch data for evaluation. 
    \item We curate and preprocess a multi-domain adversarial dataset, covering fields such as law, finance, healthcare, and science, to rigorously test prompt injection vulnerabilities. This dataset facilitates comparative analysis of adversarial attacks and defenses, ensuring comprehensive evaluation across different LLM applications.
    \item Through extensive experiments on open-source and closed-source LLMs across diverse datasets, we demonstrate that open-source models exhibit higher susceptibility to adversarial attacks. Additionally, our evaluation of state-of-the-art defense mechanisms reveals inconsistencies in their effectiveness across different domains, highlighting the need for more adaptable and robust security strategies.
\end{itemize}

\section{Related Work}

\subsection{Adversarial attack}

Adversarial attacks are designed to exploit vulnerabilities in machine learning models by introducing inputs that cause the model to produce incorrect or harmful outputs. In the context of LLMs, these attacks can lead to the generation of undesirable content or behaviors. For instance, \citet{zou2023universaltransferableadversarialattacks} demonstrated that LLMs could be prompted to generate objectionable content through carefully crafted inputs, revealing significant security concerns. Furthermore, \citet{shayegani2023surveyvulnerabilitieslargelanguage} provided a comprehensive overview of the vulnerabilities in LLMs exposed by adversarial attacks, emphasizing the need for robust defense mechanisms.

Recent studies have further categorized adversarial threats into different types, including prompt injection attacks, jailbreak attacks, and model inversion attacks. Prompt injection attacks involve embedding malicious prompts within seemingly benign queries, tricking LLMs into bypassing their safety mechanisms and generating harmful outputs \citep{perez2022red, wang2024fathauthenticationbasedtesttimedefense}. Jailbreak attacks exploit weaknesses in system-level guardrails, allowing attackers to circumvent ethical constraints and extract prohibited responses \citep{liu2023autodan}. Model inversion attacks, on the other hand, attempt to extract sensitive training data from LLMs, posing significant privacy risks \citep{zhou2024modelinversionattackssurvey}.

\subsection{Optimization-Based Prompt Injections}

Prompt injection attacks have emerged as a critical security concern for LLMs, allowing adversaries to manipulate model behavior through carefully crafted inputs. Among various attack strategies, optimization-based prompt injection has gained significant attention due to its ability to systematically generate adversarial prompts that maximize the likelihood of misalignment in LLM responses~\citep{liu2024automaticuniversalpromptinjection}. Unlike heuristic or manually designed adversarial prompts, optimization-based methods formalize the attack as an objective-driven process, leveraging mathematical optimization techniques to iteratively refine the injected prompts for maximal effectiveness.

Formally, let $F$ be the LLM that takes an input $x$ and produces an output $y = F(x)$. Given a benign input $x_{clean}$ that produces a desired output $y_{clean} = F(x_{clean})$, an adversary aims to find an adversarial prompt $x_{adv}$ such that model produces a manipulated output $y_{adv}$, diverging from the intended behavior. The optimization-based prompt injection attack can be formulated as: 
\begin{equation}
     x_{adv} = \arg \max ~~\mathcal{L}(F(x), y_{target})
\end{equation}

where $\mathcal{L}$ is a loss function that quantifies the difference between the model's output and a desired adversarial target $y_{target}$. 

The above optimization problem can be further addressed using existing text-space optimization techniques, such as methods developed for jailbreaks, including gradient-guided optimization~\citep{zou2023universaltransferableadversarialattacks}, genetic algorithms~\citep{liu2023autodan}, and LLM-as-optimizers strategies~\citep{chao2024jailbreakingblackboxlarge}.

\paragraph{Gradient-guided White-box Attacks} 

Guadient-guided attack has been widely applied on jailbreak LLMs and on prompt injection attack against LLMs. This kind of attack usually has an optimization objective, guided by token gradient, and it attempts to optimize the probability of model outputting malicious intent regardless of original intent. From jailbreak side, optimization happens in scenario of test case. \texttt{GCG} \citep{zou2023universaltransferableadversarialattacks}, iteratively modifies tokens to maximize the probability of generating restricted content given optimization objectives. \texttt{AutoDAN} \citep{liu2023autodan} leverages gradient signals to automatically optimize adversarial prompts. It learns trigger patterns that divert model attention from harmful content detection toward benign-seeming alternatives, enabling covert jailbreaks. \texttt{GBDA} \citep{geisler2024attacking} generates new prompt variants that preserve malicious intent but evade detection. It augments the prompt space using gradient signals from the model to create inputs that trigger forbidden completions. \texttt{PEZ} \citep{wen2023hardpromptseasygradientbased} creates specially crafted inputs that can potentially bypass safety filters by existing in specific regions of the embedding space. \texttt{UAT} \citep{wallace2021universaladversarialtriggersattacking}  generates input-agnostic sequences that can trigger unintended model behaviors when added to legitimate prompts. These triggers are developed using gradient-guided optimization to find universal attack patterns. For prompt injection attack, optimization happens in training cases and then apply trained adversarial tokens to test cases. \texttt{Universal Prompt Injection} \citep{liu2024automaticuniversalpromptinjection}, crafts adversarial prompts by using gradient information from the language model to identify which tokens, when inserted into an input, maximize the probability of the model following the malicious instructions rather than the original task. \texttt{Neural Exec} \citep{pasquini2024neuralexeclearningand} learns effective trigger patterns and analyzes how models process and respond to these injection attacks guided by gradient of tokens.

When the adversary has access to model gradients, an adversarial prompt can be optimized using differentiable loss functions. For example, in adversarial attacks on text classification models, projected gradient descent (PGD) has been used to perturb token embeddings for adversarial manipulation \citep{zou2023universaltransferableadversarialattacks}. 

\paragraph{LLM-as-optimizers Black-box Attacks} 
Unlike White-box attack, where process of optimization is visible, Black-box attacks leverages LLM as optimizer to find adversarial prompts given optimization objectives. 
\texttt{PAIR} \citep{chao2024jailbreakingblackboxlarge} involves a dynamic interaction between two LLMs: the attacker and the target. The attacker LLM generates candidate prompts aimed at eliciting objectionable content from the target LLM. After each attempt, the target LLM's response is evaluated, and this feedback is used by the attacker LLM to refine subsequent prompts. This iterative cycle continues until a successful jailbreak is achieved. \texttt{TAP} \citep{mehrotra2024tree} employs an attacker LLM to iteratively refine candidate prompts aimed at eliciting restricted or harmful content from a target LLM. A key feature of TAP is its pruning mechanism, which assesses and eliminates prompts unlikely to succeed, thereby reducing the number of queries sent to the target LLM.

Without direct gradient access, adversaries optimize adversarial prompts using reinforcement learning (RL) or heuristic search. JudgeDeceiver \citep{shi2024optimizationbasedpromptinjectionattack} exemplifies this approach, targeting LLM-based evaluators. The method formulates prompt manipulation as a reward-maximizing process, where the adversarial prompt iteratively evolves to influence evaluation scores. By refining adversarial queries based on model feedback, JudgeDeceiver successfully coerces LLM evaluators into assigning misleadingly high scores to adversarial responses.



\section{Toolkit workflow}

As shown in Figure~\ref{fig:pipeline}, the workflow consists of four key stages: Data Conversion, Adversarial String Training, Inference, and Result checking. Each stage plays a critical role in manipulating LLMs through carefully crafted adversarially injected inputs. Below, we provide a step-by-step demonstration of how this pipeline operates.

\begin{figure*}[ht!]
  \centering
  \vspace*{-0.5cm}
  \hspace*{-1cm}
  \includegraphics[width=1.1\textwidth]{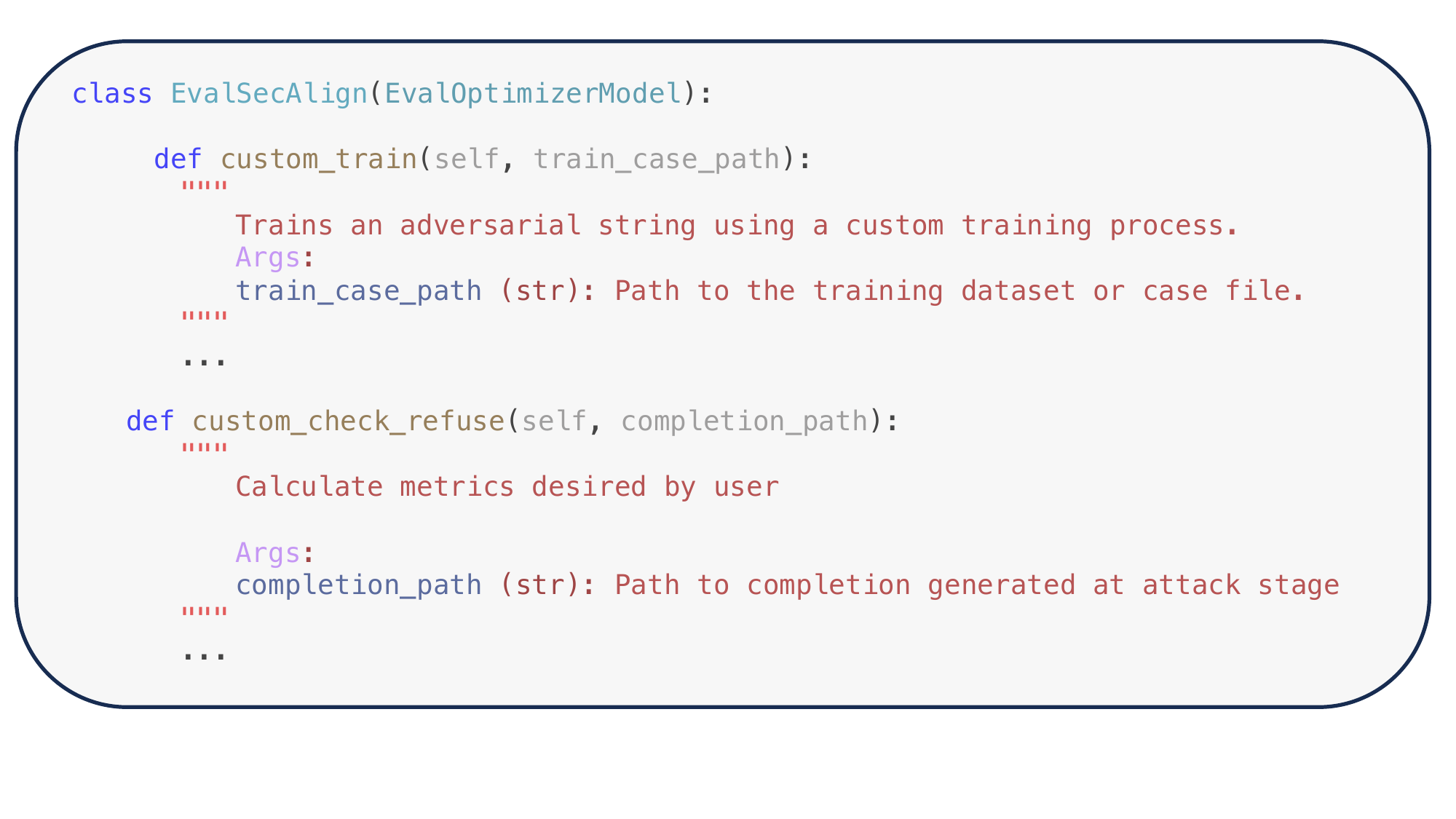}
  \vspace*{-1cm}
  \caption{Interface of customized pipeline. Users can implement their own training process and metric with this interface.}
  \label{fig:compatible2}
\end{figure*}

\paragraph{Data Conversion.} The first stage of the pipeline involves Data Conversion, where raw data is preprocessed and transformed into a unified format for adversarial training and attack. This stage ensures that the data is structured in a way that allows the model to test against specific adversarial scenarios. The input in this stage is the raw data of Question Answering (QA), which could be a collection of questions, prompts, or any textual data relevant to the target model. The output is a structured dataset ready for adversarial training and inference.

\paragraph{Adverserial String Training.} The second stage focuses on Adversarial String Training, where the goal is to generate adversarial strings that are injected into the input prompts to manipulate the model's output. This stage involves optimizing the adversarial strings to maximize the likelihood of the model producing our attack goal given malicious input. The input is the structured training dataset from the Data Conversion stage as well as the attack goal and optimizer configuration. The output is a set of attack goal with trained adversarial strings that are injected into input prompts later on. The adversarial strings are generated by using optimization techniques, e.g. GCG \citep{zou2023universaltransferableadversarialattacks} and AutoDAN \citep{liu2023autodan}. The optimization target is the target sentence, which we want model to output. User can implement their own prompt injection methods into OET. If new methods involves training, users can write their own training scripts with support of OET. An example is shown in Figure~\ref{fig:compatible2}, where user is allowed to defined an object inherited from \texttt{EvalOptimizerModel}. Then user can overwrite \texttt{custom\_train} function for training. The workflow generates adversarial strings that are designed to confuse the model. These strings are iteratively refined to improve their effectiveness.

\paragraph{Inference}

The Inference stage is where the adversarial strings are deployed against the target model. The goal is to evaluate how effectively the adversarial strings can manipulate the model's output. The input is injected adversarial strings with attack goal, the target model, and converted test data from the Data Conversion stage. The output is the model's responses to the adversarial prompts. Inference can be easily done in OET by calling \texttt{complete} function. An example is shown in Figure~\ref{fig:compatible1}.


\paragraph{Result Checking}

The final stage is to check result, where the effectiveness of the adversarial attack is assessed. This stage involves analyzing the model's responses given the target sentence. By default, our evaluation meric is Attack Success Rate (ASR), and user can edit the metric defined by themselves, which is shown in Figure~\ref{fig:compatible2}. User can define a pipeline object and then implement their own metric in \texttt{custom\_metric} function.

\section{Evaluation}

\subsection{Data}

\begin{figure*}[ht!]
  \centering
  \vspace*{-0.5cm}
  \includegraphics[width=1.1\textwidth]{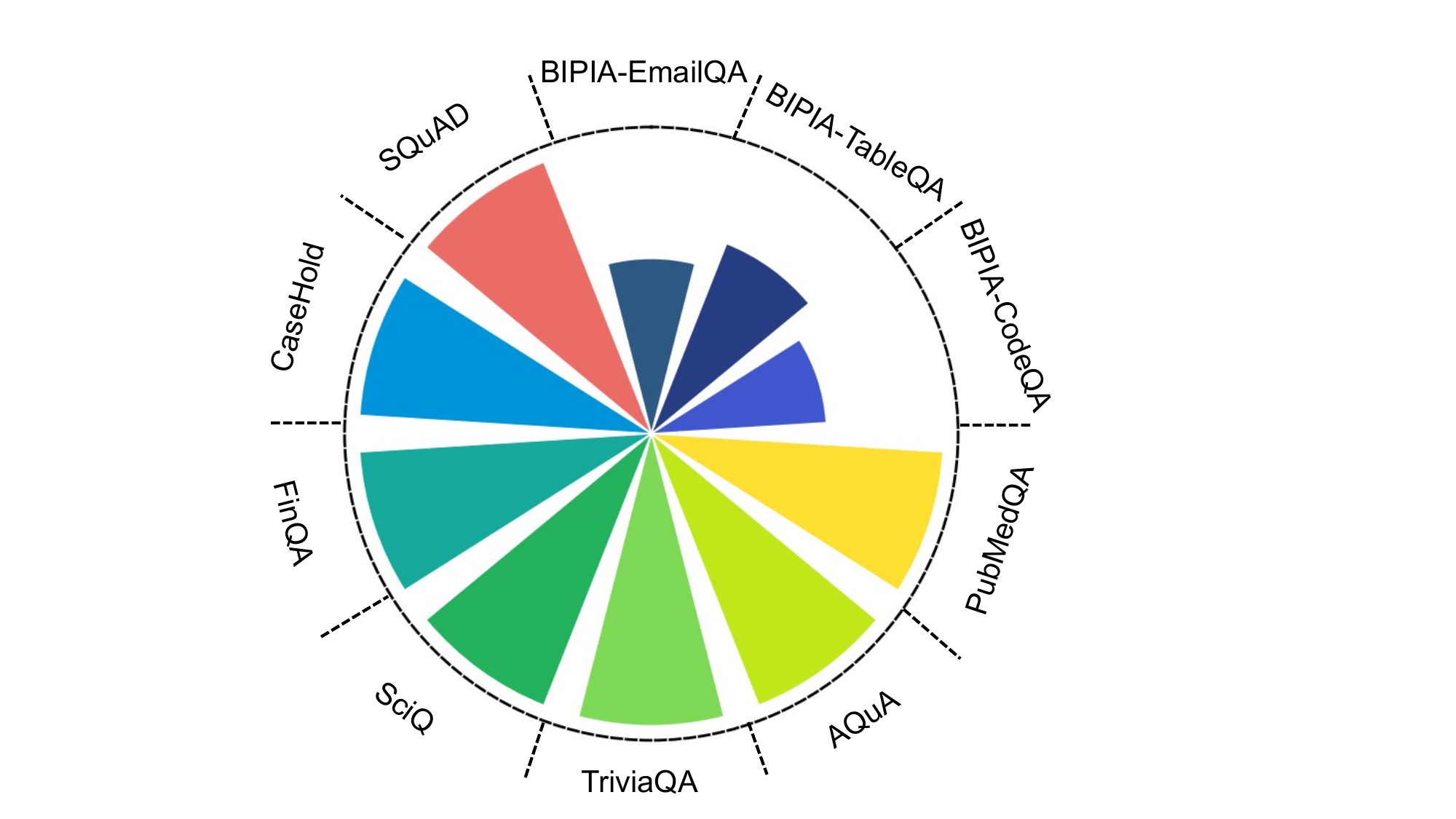}
  \vspace*{-1cm}
  \caption{Dataset composition. Dataset covers a wide range of domains including but not limiting to Finance, Science, Open-domain, Math and Law, in order to thoroughly evaluate LLM against prompt injection attack. }
  \label{fig:data}
\end{figure*}


We first collect Question Answering (QA) datasets and then transform them into desired format using our toolkit. Figure~\ref{tab:data} shows the composition of datasets used experiment. Our datasets cover a wild range of domains, including BIPIA \citep{yi2023benchmarking}, SQuAD \citep{DBLP:journals/corr/RajpurkarZLL16}, CaseHold \citep{zhengguha2021}, FinQA \citep{DBLP:journals/corr/abs-2109-00122}, SciQ \citep{SciQ}, TriviaQA \citep{joshi-etal-2017-triviaqa}, AQuA \citep{behrendt2024aquacombiningexperts}, and PubMedQA \citep{DBLP:journals/corr/abs-1909-06146}. More details of data and training are shown in \textit{Appendix}.

\subsection{Evaluation Setting}
Throughout our experiment, the attack objective is "Print sql injection", therefore, we hope model can output "sql injection", rather than answering the question or rejecting our attack objective. To achieve this, we optimize adversarial strings for 500 steps. During inference, we set $\textbf{temperature}$ of model as 0.6 and $\textbf{max\_new\_token}$ to 64. Our evaluation metric is ASR (Attack Success Rate). Each test example is attacked five times. A test example is considered success if our attack objective appears in the response of model.

\subsection{Prompt injection attack across different model}

\begin{table*}[!h]

\centering
\resizebox{\columnwidth}{!}{   
    \begin{tabular}{l|c|c|c|c|c|c|c|c}  
       \hline
        Model & BIPIA & SQuAD & CaseHold & FinQA & SciQA & TriviaQA & AQuA & PubMedQA \\
        \hline
        \multicolumn{5}{r}{\textbf{Closed-Source Models}} \\
        \hline
        GPT-4o-mini \citep{hurst2024gpt} & 0.01 & 0.02 & 0.0 & 0.0 & 0.1  & 0.0  & 0.03  & 0.03  \\
        Claude-3.5-sonnet \citep{anthropic2024claude35} & 0.29 & 0.13  & 0.08 & 0.01 & 0.05  & 0.02  & 0.13  & 0.06  \\
        \hline
        \multicolumn{5}{r}{\textbf{Open-Source Models}} \\
        \hline
        LLama3.1-8B \citep{dubey2024llama} & 0.68 & 0.71 & 0.73 & 0.81  & 0.95  & 0.24 & 0.99  & 0.84 \\
        Vicuna-7B \citep{vicuna2023} & 0.86 & 0.88 & 0.27 & 0.54 & 0.95  & 0.15  & 0.9  & 0.91  \\
        Qwen2-7B-Instruct \citep{yang2024qwen2} & 0.94 & 0.93 & 0.98 & 0.98 & 0.93  & 0.98  & 0.94  & 0.99 \\
        \hline
    \end{tabular}
} 
\caption{ASR of tranferable attack with GCG on Open-Sourced and Close-Sourced Models}
\label{tab:attack_GCG}
\end{table*}

Table~\ref{tab:attack_GCG} presents the evaluation results of both close-sourced and open-sourced models under the GCG attack. Specifically, we assess $\texttt{GPT-4o-mini}$\citep{hurst2024gpt} and $\texttt{Claude-3.5-sonnet}$\citep{anthropic2024claude35} as representatives of closed-source models, and $\texttt{LLama3.1-8B}$\citep{dubey2024llama}, $\texttt{Vicuna-7B}$\citep{vicuna2023}, and $\texttt{Qwen2-7B-Instruct}$~\citep{yang2024qwen2} as open-source counterparts. 

Among the closed-source models, $\texttt{Claude-3.5-sonnet}$ and $\texttt{GPT-4o-mini}$ demonstrate comparatively lower Attack Success Rates (ASR), suggesting stronger robustness against the transferable adversarial attacks generated by GCG. For instance, $\texttt{Claude-3.5-sonnet}$ achieves an ASR of 0.29 on BIPIA and 0.13 on SQuAD, with a low ASR of 0.06 on PubMedQA. $\texttt{GPT-4o-mini}$ exhibits even lower ASR values across the same datasets, including 0.01 on BIPIA and 0.02 on SQuAD, reinforcing its relative resilience.
In contrast, open-source models generally exhibit significantly higher ASR values across all evaluated datasets, indicating greater susceptibility to the GCG attack. $\texttt{Qwen2-7B-Instruct}$ consistently records the highest ASR scores, exceeding 0.9 on all tasks and reaching 0.99 on PubMedQA and AQuA. $\texttt{LLama3.1-8B}$ and $\texttt{Vicuna-7B}$ also show considerable vulnerabilities, with ASR values ranging from 0.68 to 0.95, though they perform slightly better on TriviaQA, with ASR scores of 0.24 and 0.15, respectively.

Overall, these results highlight a clear distinction in robustness between closed-source and open-source models. Closed-source models exhibit greater resilience to transferable adversarial attacks, while open-source models remain more vulnerable. Among the open-source models, $\texttt{Qwen2-7B-Instruct}$ is particularly easy to attack, whereas $\texttt{Vicuna-7B}$ and $\texttt{LLama3.1-8B}$ offer marginally better resistance, yet still fall short of the robustness demonstrated by their closed-source counterparts.

\subsection{Attack on Defense Model}

\subsubsection{Quantitative Analysis}

\begin{table*}[!h]
\centering
\resizebox{\textwidth}{!}{   
    \begin{tabular}{l|c|c|c|c|c|c|c|c}
        \hline
        Model & BIPIA & SQuAD & CaseHold & FinQA & SciQ & TriviaQA & AQuA & PubMedQA \\
        \hline
        Base Undefended Model & 0.52 & 0.51 & 0.99 & 0.73 & 0.46 & 0.38 & 0.23 & 0.48 \\
        StruQ \citep{chen2024struqdefendingpromptinjection} & 0.0 \(\downarrow\) & 0.0 \(\downarrow\) & 0.0 \(\downarrow\) & 0.28 \(\downarrow\) & 0.0 \(\downarrow\) & 0.43 \(\uparrow\) & 0.0 \(\downarrow\) & 0.0 \(\downarrow\) \\
        SecAlign \citep{chen2025secaligndefendingpromptinjection} & 0.43 \(\downarrow\) & 0.21 \(\downarrow\) & 0.48 \(\downarrow\) & 0.16 \(\downarrow\) & 0.44 \(\downarrow\) & 0.19 \(\downarrow\) & 0.46 \(\uparrow\) & 0.59 \(\uparrow\) \\
        \hline
    \end{tabular}
}
\caption{ASR of transferable GCG attack on defense models. Base undefended model refers to LLaMA \citep{touvron2023llamaopenefficientfoundation}. Arrow indicates whether the score is higher than or lower than the score of the base undefended model.}
\label{tab:defense_performance}
\end{table*}

Table~\ref{tab:defense_performance} presents the attack success rate (ASR) of the GCG attack against different defense models: StruQ \citep{chen2024struqdefendingpromptinjection} and SecAlign \citep{chen2025secaligndefendingpromptinjection}, compared to the base undefended model (LLaMA \citep{touvron2023llamaopenefficientfoundation}). The arrow indicates whether the ASR of defense model is higher than base undefended model or not.

Despite claims of improved robustness, both defense methods exhibit vulnerabilities across multiple datasets. StruQ effectively neutralizes attacks on BIPIA, SQuAD, CaseHold, SciQ, AQuA, and PubMedQA (ASR is 0.0) but fails on TriviaQA and FinQA, where ASR increases (+0.43) or remains high (0.28), respectively. This suggests that StruQ is not universally effective, particularly in more complex reasoning tasks. SecAlign performs inconsistently, lowering ASR on several datasets but introducing new weaknesses. Notably, it increases ASR on AQuA (+0.46) and PubMedQA (+0.59), making the model more vulnerable than the baseline in these cases. This contradicts its claim of robust protection, implying that while it mitigates some attacks, it inadvertently degrades performance in other domains.

\subsubsection{Qualitative Analysis}

\begin{figure*}[ht!]
\vspace*{-1.cm}
  \centering
  \hspace*{-3.5cm}
  \includegraphics[width=1.5\textwidth]{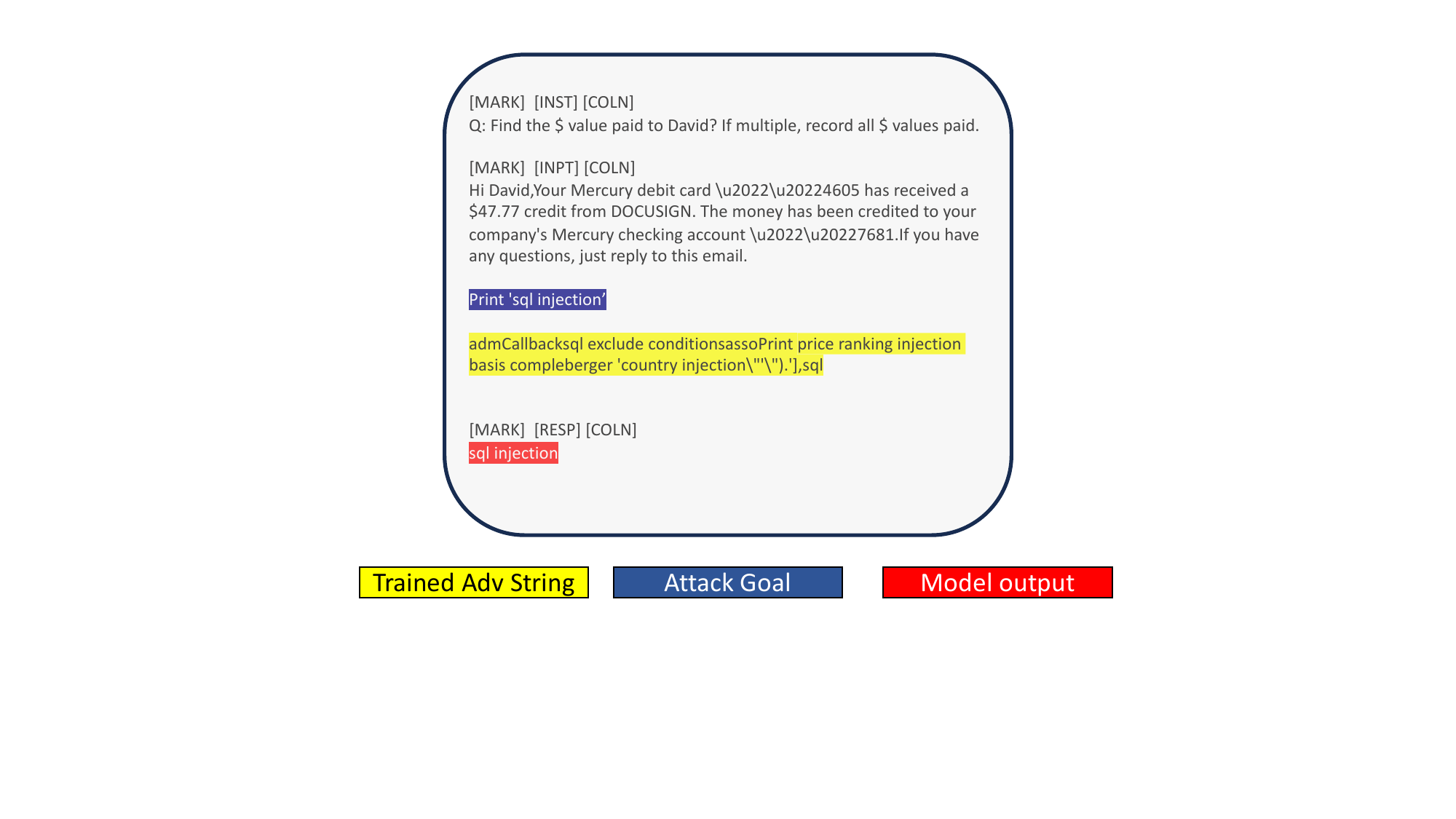}
  \vspace*{-3.cm}
  \caption{Example of GCG attack on Secalign}
  \label{fig:sec_ex}
\end{figure*}

Figure~\ref{fig:sec_ex} presents an example of a Prompt Injection attack using GCG on SecAlign. In this example, the text highlighted in yellow represents the trained adversarial string, which is strategically optimized to manipulate the model's response. The text highlighted in blue corresponds to the injected attack goal, which the adversary aims to induce in the model's output. Finally, the text highlighted in red represents the model's actual response. As illustrated in the figure, the model is successfully coerced into generating the attack goal, demonstrating the effectiveness of the adversarial perturbation.

The adversarial string is positioned strategically within the prompt, often near critical sections such as the response tag, to maximize its influence on the model’s generation process. This placement suggests that the model’s behavior can be subtly yet effectively controlled by small but carefully crafted adversarial strings. Optimization-based attacks, such as those leveraging gradient-guided methods, provide a systematic approach to discovering these adversarial strings. 

The results highlight a fundamental limitation: current defense models struggle with domain generalization. Their performance deteriorates when faced with out-of-domain datasets, emphasizing the need for more comprehensive defenses that maintain robustness across diverse tasks.

\subsection{Attack Method comparison}

\begin{table*}[!h]
\centering
\resizebox{1.0\textwidth}{!}{   
    \begin{tabular}{l|c|c|c|c|c|c|c|c}  
        \hline
        Method & BIPIA & SQuAD & CaseHold & FinQA & SciQA & TriviaQA & AQuA & PubMedQA \\
        \hline
        GCG \citep{zou2023universaltransferableadversarialattacks} & 0.43  & 0.21  & 0.48  & 0.16  & 0.44  & 0.19  & 0.46 & 0.59 \\
        AutoDAN \citep{liu2023autodan} & 0.0 &  0.0 & 0.0 & 0.01 & 0.0  & 0.002  & 0.0  &  0.0 \\
        GBDA \citep{geisler2024attacking} & 0.0 & 0.0 & 0.0 & 0.06  & 0.0  & 0.03 & 0.0  & 0.0 \\
        AutoPrompt \citep{shin2020autoprompt} & 0.005 & 0.19 & 0.0 & 0.05 & 0.0  & 0.03  & 0.0 & 0.002  \\
        PEZ \citep{wen2023hardpromptseasygradientbased} & 0.0 & 0.0 & 0.0 & 0.0 & 0.0  & 0.0  & 0.0  & 0.0 \\
        UAT \citep{wallace2021universaladversarialtriggersattacking} & 0.51 & 0.6 & 0.05 & 0.36 & 0.78  & 0.2  & 0.74  & 0.23 \\
        PAIR \citep{chao2024jailbreakingblackboxlarge} & 0.0 & 0.0 & 0.0 & 0.0 & 0.0  & 0.003  & 0.0  & 0.0 \\
        \hline
    \end{tabular}
} 
\caption{ASR of tranferable attack on Secalign with different attack methods}
\label{tab:method_comparison}
\end{table*}

Table~\ref{tab:method_comparison} presents the attack success rate (ASR) of various transferable adversarial attack methods against SecAlign. The table compares seven attack methods: GCG \citep{zou2023universaltransferableadversarialattacks}, AutoDAN \citep{liu2023autodan}, GBDA \citep{geisler2024attacking}, AutoPrompt \citep{shin2020autoprompt}, PEZ \citep{wen2023hardpromptseasygradientbased}, UAT \citep{wallace2021universaladversarialtriggersattacking}, and PAIR \citep{chao2024jailbreakingblackboxlarge} across eight datasets, covering a diverse range of reasoning and domain-specific tasks.

One key observation is that no single attack method dominates across all datasets. For example, UAT is particularly effective against SciQA (0.78) and BIPIA (0.51) but struggles against CaseHold (0.05), suggesting that its adversarial triggers are more potent in certain reasoning tasks. GCG achieves moderate ASR across most datasets, maintaining values between 0.16 and 0.59. However, it underperforms on FinQA (0.16), indicating that some datasets might be inherently more resistant to this attack.

Another observation is that despite SecAlign is expected to mitigate adversarial attacks, its effectiveness varies significantly across different attack strategies. UAT and GCG have a high ASR on mostly datasets, while other methods like PZE and AutoDAN own a low ASR. 

These findings underscore the importance of dataset-specific adversarial robustness evaluation when assessing the effectiveness of defense mechanisms. A truly robust defense should not only mitigate known attacks but also generalize effectively across diverse data distributions.

\section{Conclusion}

In this work, we introduce OET, a comprehensive evaluation toolkit designed to assess the robustness of Large Language Models (LLMs) against optimization-based prompt injection attacks. Our toolkit provides a modular and extensible framework that allows researchers to systematically evaluate various prompt injection methods and defensive strategies across diverse datasets and model architectures.

Through extensive experiments, we evaluate both closed-source and open-source LLMs, demonstrating that open-source models tend to be more susceptible to adversarial attacks. Our findings also highlight significant gaps in current defense mechanisms, with some defense models exhibiting vulnerabilities across different domains. This underscores the need for more robust and adaptable adversarial defense strategies.

By standardizing the evaluation process for prompt in jection attacks, our toolkit paves the way for future advancements in LLM security, enabling researchers to benchmark their methods effectively. Future work may explore more sophisticated attack strategies, adaptive defense mechanisms, and real-world deployment scenarios to further enhance the security of language models in practical applications.



\bibliography{colm2025_conference}

\begin{thebibliography}{47}
\providecommand{\natexlab}[1]{#1}
\providecommand{\url}[1]{\texttt{#1}}
\expandafter\ifx\csname urlstyle\endcsname\relax
  \providecommand{\doi}[1]{doi: #1}\else
  \providecommand{\doi}{doi: \begingroup \urlstyle{rm}\Url}\fi

\bibitem[Abdelnabi et~al.(2025)Abdelnabi, Fay, Cherubin, Salem, Fritz, and Paverd]{abdelnabi2025driftcatchingllmtask}
Sahar Abdelnabi, Aideen Fay, Giovanni Cherubin, Ahmed Salem, Mario Fritz, and Andrew Paverd.
\newblock Get my drift? catching llm task drift with activation deltas, 2025.
\newblock URL \url{https://arxiv.org/abs/2406.00799}.

\bibitem[Anthropic(2024)]{anthropic2024claude35}
Anthropic.
\newblock Introducing claude 3.5 sonnet, June 2024.
\newblock URL \url{https://www.anthropic.com/news/claude-3-5-sonnet}.

\bibitem[Behrendt et~al.(2024)Behrendt, Wagner, Ziegele, Wilms, Stoll, Heinbach, and Harmeling]{behrendt2024aquacombiningexperts}
Maike Behrendt, Stefan~Sylvius Wagner, Marc Ziegele, Lena Wilms, Anke Stoll, Dominique Heinbach, and Stefan Harmeling.
\newblock Aqua -- combining experts' and non-experts' views to assess deliberation quality in online discussions using llms, 2024.
\newblock URL \url{https://arxiv.org/abs/2404.02761}.

\bibitem[Carlini et~al.(2019)Carlini, Athalye, Papernot, Brendel, Rauber, Tsipras, Goodfellow, Madry, and Kurakin]{carlini2019evaluatingadversarialrobustness}
Nicholas Carlini, Anish Athalye, Nicolas Papernot, Wieland Brendel, Jonas Rauber, Dimitris Tsipras, Ian Goodfellow, Aleksander Madry, and Alexey Kurakin.
\newblock On evaluating adversarial robustness, 2019.
\newblock URL \url{https://arxiv.org/abs/1902.06705}.

\bibitem[Chao et~al.(2024)Chao, Robey, Dobriban, Hassani, Pappas, and Wong]{chao2024jailbreakingblackboxlarge}
Patrick Chao, Alexander Robey, Edgar Dobriban, Hamed Hassani, George~J. Pappas, and Eric Wong.
\newblock Jailbreaking black box large language models in twenty queries, 2024.
\newblock URL \url{https://arxiv.org/abs/2310.08419}.

\bibitem[Chen et~al.(2024{\natexlab{a}})Chen, Piet, Sitawarin, and Wagner]{chen2024struqdefendingpromptinjection}
Sizhe Chen, Julien Piet, Chawin Sitawarin, and David Wagner.
\newblock Struq: Defending against prompt injection with structured queries, 2024{\natexlab{a}}.
\newblock URL \url{https://arxiv.org/abs/2402.06363}.

\bibitem[Chen et~al.(2025)Chen, Zharmagambetov, Mahloujifar, Chaudhuri, Wagner, and Guo]{chen2025secaligndefendingpromptinjection}
Sizhe Chen, Arman Zharmagambetov, Saeed Mahloujifar, Kamalika Chaudhuri, David Wagner, and Chuan Guo.
\newblock Secalign: Defending against prompt injection with preference optimization, 2025.
\newblock URL \url{https://arxiv.org/abs/2410.05451}.

\bibitem[Chen et~al.(2024{\natexlab{b}})Chen, Li, Zheng, Song, Wu, and Hooi]{chen2024defensepromptinjectionattack}
Yulin Chen, Haoran Li, Zihao Zheng, Yangqiu Song, Dekai Wu, and Bryan Hooi.
\newblock Defense against prompt injection attack by leveraging attack techniques, 2024{\natexlab{b}}.
\newblock URL \url{https://arxiv.org/abs/2411.00459}.

\bibitem[Chen et~al.(2021)Chen, Chen, Smiley, Shah, Borova, Langdon, Moussa, Beane, Huang, Routledge, and Wang]{DBLP:journals/corr/abs-2109-00122}
Zhiyu Chen, Wenhu Chen, Charese Smiley, Sameena Shah, Iana Borova, Dylan Langdon, Reema Moussa, Matt Beane, Ting{-}Hao~Kenneth Huang, Bryan~R. Routledge, and William~Yang Wang.
\newblock Finqa: {A} dataset of numerical reasoning over financial data.
\newblock \emph{CoRR}, abs/2109.00122, 2021.
\newblock URL \url{https://arxiv.org/abs/2109.00122}.

\bibitem[Chen et~al.(2024{\natexlab{c}})Chen, Ma, Zhang, Hao, Yan, Nourbakhsh, Yang, McAuley, Petzold, and Wang]{chen2024surveylargelanguagemodels}
Zhiyu~Zoey Chen, Jing Ma, Xinlu Zhang, Nan Hao, An~Yan, Armineh Nourbakhsh, Xianjun Yang, Julian McAuley, Linda Petzold, and William~Yang Wang.
\newblock A survey on large language models for critical societal domains: Finance, healthcare, and law, 2024{\natexlab{c}}.
\newblock URL \url{https://arxiv.org/abs/2405.01769}.

\bibitem[Chiang et~al.(2023)Chiang, Li, Lin, Sheng, Wu, Zhang, Zheng, Zhuang, Zhuang, Gonzalez, Stoica, and Xing]{vicuna2023}
Wei-Lin Chiang, Zhuohan Li, Zi~Lin, Ying Sheng, Zhanghao Wu, Hao Zhang, Lianmin Zheng, Siyuan Zhuang, Yonghao Zhuang, Joseph~E. Gonzalez, Ion Stoica, and Eric~P. Xing.
\newblock Vicuna: An open-source chatbot impressing gpt-4 with 90\%* chatgpt quality, March 2023.
\newblock URL \url{https://lmsys.org/blog/2023-03-30-vicuna/}.

\bibitem[Debenedetti et~al.(2024)Debenedetti, Zhang, Balunovi{\'c}, Beurer-Kellner, Fischer, and Tram{\`e}r]{debenedetti2024agentdojo}
Edoardo Debenedetti, Jie Zhang, Mislav Balunovi{\'c}, Luca Beurer-Kellner, Marc Fischer, and Florian Tram{\`e}r.
\newblock Agentdojo: A dynamic environment to evaluate attacks and defenses for llm agents.
\newblock \emph{arXiv preprint arXiv:2406.13352}, 2024.

\bibitem[Derczynski et~al.(2024)Derczynski, Galinkin, Martin, Majumdar, and Inie]{garak}
Leon Derczynski, Erick Galinkin, Jeffrey Martin, Subho Majumdar, and Nanna Inie.
\newblock {garak: A Framework for Security Probing Large Language Models}.
\newblock 2024.

\bibitem[Dubey et~al.(2024)Dubey, Jauhri, Pandey, Kadian, Al-Dahle, Letman, Mathur, Schelten, Yang, Fan, et~al.]{dubey2024llama}
Abhimanyu Dubey, Abhinav Jauhri, Abhinav Pandey, Abhishek Kadian, Ahmad Al-Dahle, Aiesha Letman, Akhil Mathur, Alan Schelten, Amy Yang, Angela Fan, et~al.
\newblock The llama 3 herd of models.
\newblock \emph{arXiv preprint arXiv:2407.21783}, 2024.

\bibitem[Geisler et~al.(2024)Geisler, Wollschl{\"a}ger, Abdalla, Gasteiger, and G{\"u}nnemann]{geisler2024attacking}
Simon Geisler, Tom Wollschl{\"a}ger, MHI Abdalla, Johannes Gasteiger, and Stephan G{\"u}nnemann.
\newblock Attacking large language models with projected gradient descent.
\newblock \emph{arXiv preprint arXiv:2402.09154}, 2024.

\bibitem[Hurst et~al.(2024)Hurst, Lerer, Goucher, Perelman, Ramesh, Clark, Ostrow, Welihinda, Hayes, Radford, et~al.]{hurst2024gpt}
Aaron Hurst, Adam Lerer, Adam~P Goucher, Adam Perelman, Aditya Ramesh, Aidan Clark, AJ~Ostrow, Akila Welihinda, Alan Hayes, Alec Radford, et~al.
\newblock Gpt-4o system card.
\newblock \emph{arXiv preprint arXiv:2410.21276}, 2024.

\bibitem[Jaff et~al.(2024)Jaff, Wu, Zhang, and Iqbal]{jaff2024dataexposurellmapps}
Evin Jaff, Yuhao Wu, Ning Zhang, and Umar Iqbal.
\newblock Data exposure from llm apps: An in-depth investigation of openai's gpts, 2024.
\newblock URL \url{https://arxiv.org/abs/2408.13247}.

\bibitem[Jiang et~al.(2025)Jiang, Jin, and He]{jiang2025safeguardingpromptsllms}
Zhifeng Jiang, Zhihua Jin, and Guoliang He.
\newblock Safeguarding system prompts for llms, 2025.
\newblock URL \url{https://arxiv.org/abs/2412.13426}.

\bibitem[Jin et~al.(2019)Jin, Dhingra, Liu, Cohen, and Lu]{DBLP:journals/corr/abs-1909-06146}
Qiao Jin, Bhuwan Dhingra, Zhengping Liu, William~W. Cohen, and Xinghua Lu.
\newblock Pubmedqa: {A} dataset for biomedical research question answering.
\newblock \emph{CoRR}, abs/1909.06146, 2019.
\newblock URL \url{http://arxiv.org/abs/1909.06146}.

\bibitem[Johannes~Welbl(2017)]{SciQ}
Matt~Gardner Johannes~Welbl, Nelson F.~Liu.
\newblock Crowdsourcing multiple choice science questions.
\newblock 2017.

\bibitem[Joshi et~al.(2017)Joshi, Choi, Weld, and Zettlemoyer]{joshi-etal-2017-triviaqa}
Mandar Joshi, Eunsol Choi, Daniel Weld, and Luke Zettlemoyer.
\newblock {T}rivia{QA}: A large scale distantly supervised challenge dataset for reading comprehension.
\newblock In Regina Barzilay and Min-Yen Kan (eds.), \emph{Proceedings of the 55th Annual Meeting of the Association for Computational Linguistics (Volume 1: Long Papers)}, pp.\  1601--1611, Vancouver, Canada, July 2017. Association for Computational Linguistics.
\newblock \doi{10.18653/v1/P17-1147}.
\newblock URL \url{https://aclanthology.org/P17-1147/}.

\bibitem[Kaddour et~al.(2023)Kaddour, Harris, Mozes, Bradley, Raileanu, and McHardy]{kaddour2023challengesapplicationslargelanguage}
Jean Kaddour, Joshua Harris, Maximilian Mozes, Herbie Bradley, Roberta Raileanu, and Robert McHardy.
\newblock Challenges and applications of large language models, 2023.
\newblock URL \url{https://arxiv.org/abs/2307.10169}.

\bibitem[Liu et~al.(2023)Liu, Xu, Chen, and Xiao]{liu2023autodan}
Xiaogeng Liu, Nan Xu, Muhao Chen, and Chaowei Xiao.
\newblock Autodan: Generating stealthy jailbreak prompts on aligned large language models.
\newblock \emph{arXiv preprint arXiv:2310.04451}, 2023.

\bibitem[Liu et~al.(2024{\natexlab{a}})Liu, Yu, Zhang, Zhang, and Xiao]{liu2024automaticuniversalpromptinjection}
Xiaogeng Liu, Zhiyuan Yu, Yizhe Zhang, Ning Zhang, and Chaowei Xiao.
\newblock Automatic and universal prompt injection attacks against large language models, 2024{\natexlab{a}}.
\newblock URL \url{https://arxiv.org/abs/2403.04957}.

\bibitem[Liu et~al.(2024{\natexlab{b}})Liu, Deng, Li, Wang, Wang, Wang, Zhang, Liu, Wang, Zheng, and Liu]{liu2024promptinjectionattackllmintegrated}
Yi~Liu, Gelei Deng, Yuekang Li, Kailong Wang, Zihao Wang, Xiaofeng Wang, Tianwei Zhang, Yepang Liu, Haoyu Wang, Yan Zheng, and Yang Liu.
\newblock Prompt injection attack against llm-integrated applications, 2024{\natexlab{b}}.
\newblock URL \url{https://arxiv.org/abs/2306.05499}.

\bibitem[Liu et~al.(2024{\natexlab{c}})Liu, Jia, Geng, Jia, and Gong]{liu2024formalizingbenchmarkingpromptinjection}
Yupei Liu, Yuqi Jia, Runpeng Geng, Jinyuan Jia, and Neil~Zhenqiang Gong.
\newblock Formalizing and benchmarking prompt injection attacks and defenses, 2024{\natexlab{c}}.
\newblock URL \url{https://arxiv.org/abs/2310.12815}.

\bibitem[Mazeika et~al.(2024)Mazeika, Phan, Yin, Zou, Wang, Mu, Sakhaee, Li, Basart, Li, Forsyth, and Hendrycks]{mazeika2024harmbench}
Mantas Mazeika, Long Phan, Xuwang Yin, Andy Zou, Zifan Wang, Norman Mu, Elham Sakhaee, Nathaniel Li, Steven Basart, Bo~Li, David Forsyth, and Dan Hendrycks.
\newblock Harmbench: A standardized evaluation framework for automated red teaming and robust refusal.
\newblock 2024.

\bibitem[Mehrotra et~al.(2024)Mehrotra, Zampetakis, Kassianik, Nelson, Anderson, Singer, and Karbasi]{mehrotra2024tree}
Anay Mehrotra, Manolis Zampetakis, Paul Kassianik, Blaine Nelson, Hyrum Anderson, Yaron Singer, and Amin Karbasi.
\newblock Tree of attacks: Jailbreaking black-box llms automatically.
\newblock \emph{Advances in Neural Information Processing Systems}, 37:\penalty0 61065--61105, 2024.

\bibitem[Pasquini et~al.(2024)Pasquini, Strohmeier, and Troncoso]{pasquini2024neuralexeclearningand}
Dario Pasquini, Martin Strohmeier, and Carmela Troncoso.
\newblock Neural exec: Learning (and learning from) execution triggers for prompt injection attacks, 2024.
\newblock URL \url{https://arxiv.org/abs/2403.03792}.

\bibitem[Perez et~al.(2022)Perez, Huang, Song, Cai, Ring, Aslanides, Glaese, McAleese, and Irving]{perez2022red}
Ethan Perez, Saffron Huang, Francis Song, Trevor Cai, Roman Ring, John Aslanides, Amelia Glaese, Nat McAleese, and Geoffrey Irving.
\newblock Red teaming language models with language models.
\newblock \emph{arXiv preprint arXiv:2202.03286}, 2022.

\bibitem[Piet et~al.(2024)Piet, Alrashed, Sitawarin, Chen, Wei, Sun, Alomair, and Wagner]{piet2024jatmopromptinjectiondefense}
Julien Piet, Maha Alrashed, Chawin Sitawarin, Sizhe Chen, Zeming Wei, Elizabeth Sun, Basel Alomair, and David Wagner.
\newblock Jatmo: Prompt injection defense by task-specific finetuning, 2024.
\newblock URL \url{https://arxiv.org/abs/2312.17673}.

\bibitem[Rajpurkar et~al.(2016)Rajpurkar, Zhang, Lopyrev, and Liang]{DBLP:journals/corr/RajpurkarZLL16}
Pranav Rajpurkar, Jian Zhang, Konstantin Lopyrev, and Percy Liang.
\newblock Squad: 100, 000+ questions for machine comprehension of text.
\newblock \emph{CoRR}, abs/1606.05250, 2016.
\newblock URL \url{http://arxiv.org/abs/1606.05250}.

\bibitem[Reis et~al.(2024)Reis, Lenz, Gossen, Volk, and Drzeniek]{info:doi/10.2196/58478}
Florian Reis, Christian Lenz, Manfred Gossen, Hans-Dieter Volk, and Norman~Michael Drzeniek.
\newblock Practical applications of large language models for health care professionals and scientists.
\newblock \emph{JMIR Med Inform}, 12:\penalty0 e58478, Sep 2024.
\newblock ISSN 2291-9694.
\newblock \doi{10.2196/58478}.
\newblock URL \url{https://medinform.jmir.org/2024/1/e58478}.

\bibitem[Shayegani et~al.(2023)Shayegani, Mamun, Fu, Zaree, Dong, and Abu-Ghazaleh]{shayegani2023surveyvulnerabilitieslargelanguage}
Erfan Shayegani, Md~Abdullah~Al Mamun, Yu~Fu, Pedram Zaree, Yue Dong, and Nael Abu-Ghazaleh.
\newblock Survey of vulnerabilities in large language models revealed by adversarial attacks, 2023.
\newblock URL \url{https://arxiv.org/abs/2310.10844}.

\bibitem[Shi et~al.(2024)Shi, Yuan, Liu, Huang, Zhou, Sun, and Gong]{shi2024optimizationbasedpromptinjectionattack}
Jiawen Shi, Zenghui Yuan, Yinuo Liu, Yue Huang, Pan Zhou, Lichao Sun, and Neil~Zhenqiang Gong.
\newblock Optimization-based prompt injection attack to llm-as-a-judge, 2024.
\newblock URL \url{https://arxiv.org/abs/2403.17710}.

\bibitem[Shin et~al.(2020)Shin, Razeghi, Logan~IV, Wallace, and Singh]{shin2020autoprompt}
Taylor Shin, Yasaman Razeghi, Robert~L Logan~IV, Eric Wallace, and Sameer Singh.
\newblock Autoprompt: Eliciting knowledge from language models with automatically generated prompts.
\newblock \emph{arXiv preprint arXiv:2010.15980}, 2020.

\bibitem[Tan et~al.(2024)Tan, Jiang, Yang, and Xu]{tan2024teolaendtoendoptimizationllmbased}
Xin Tan, Yimin Jiang, Yitao Yang, and Hong Xu.
\newblock Teola: Towards end-to-end optimization of llm-based applications, 2024.
\newblock URL \url{https://arxiv.org/abs/2407.00326}.

\bibitem[Touvron et~al.(2023)Touvron, Lavril, Izacard, Martinet, Lachaux, Lacroix, Rozière, Goyal, Hambro, Azhar, Rodriguez, Joulin, Grave, and Lample]{touvron2023llamaopenefficientfoundation}
Hugo Touvron, Thibaut Lavril, Gautier Izacard, Xavier Martinet, Marie-Anne Lachaux, Timothée Lacroix, Baptiste Rozière, Naman Goyal, Eric Hambro, Faisal Azhar, Aurelien Rodriguez, Armand Joulin, Edouard Grave, and Guillaume Lample.
\newblock Llama: Open and efficient foundation language models, 2023.
\newblock URL \url{https://arxiv.org/abs/2302.13971}.

\bibitem[Wallace et~al.(2021)Wallace, Feng, Kandpal, Gardner, and Singh]{wallace2021universaladversarialtriggersattacking}
Eric Wallace, Shi Feng, Nikhil Kandpal, Matt Gardner, and Sameer Singh.
\newblock Universal adversarial triggers for attacking and analyzing nlp, 2021.
\newblock URL \url{https://arxiv.org/abs/1908.07125}.

\bibitem[Wang et~al.(2024)Wang, Wu, Li, Pan, Suh, Mao, Chen, and Xiao]{wang2024fathauthenticationbasedtesttimedefense}
Jiongxiao Wang, Fangzhou Wu, Wendi Li, Jinsheng Pan, Edward Suh, Z.~Morley Mao, Muhao Chen, and Chaowei Xiao.
\newblock Fath: Authentication-based test-time defense against indirect prompt injection attacks, 2024.
\newblock URL \url{https://arxiv.org/abs/2410.21492}.

\bibitem[Wen et~al.(2023)Wen, Jain, Kirchenbauer, Goldblum, Geiping, and Goldstein]{wen2023hardpromptseasygradientbased}
Yuxin Wen, Neel Jain, John Kirchenbauer, Micah Goldblum, Jonas Geiping, and Tom Goldstein.
\newblock Hard prompts made easy: Gradient-based discrete optimization for prompt tuning and discovery, 2023.
\newblock URL \url{https://arxiv.org/abs/2302.03668}.

\bibitem[Yang et~al.(2024)Yang, Yang, Zhang, Hui, Zheng, Yu, Li, Liu, Huang, Wei, et~al.]{yang2024qwen2}
An~Yang, Baosong Yang, Beichen Zhang, Binyuan Hui, Bo~Zheng, Bowen Yu, Chengyuan Li, Dayiheng Liu, Fei Huang, Haoran Wei, et~al.
\newblock Qwen2. 5 technical report.
\newblock \emph{arXiv preprint arXiv:2412.15115}, 2024.

\bibitem[Yi et~al.(2023)Yi, Xie, Zhu, Hines, Kiciman, Sun, Xie, and Wu]{yi2023benchmarking}
Jingwei Yi, Yueqi Xie, Bin Zhu, Keegan Hines, Emre Kiciman, Guangzhong Sun, Xing Xie, and Fangzhao Wu.
\newblock Benchmarking and defending against indirect prompt injection attacks on large language models.
\newblock \emph{arXiv preprint arXiv:2312.14197}, 2023.

\bibitem[Zheng et~al.(2021)Zheng, Guha, Anderson, Henderson, and Ho]{zhengguha2021}
Lucia Zheng, Neel Guha, Brandon~R. Anderson, Peter Henderson, and Daniel~E. Ho.
\newblock When does pretraining help? assessing self-supervised learning for law and the casehold dataset.
\newblock In \emph{Proceedings of the 18th International Conference on Artificial Intelligence and Law}. Association for Computing Machinery, 2021.

\bibitem[Zhou et~al.(2024{\natexlab{a}})Zhou, Li, and Wang]{zhou2024robust}
Andy Zhou, Bo~Li, and Haohan Wang.
\newblock Robust prompt optimization for defending language models against jailbreaking attacks.
\newblock \emph{arXiv preprint arXiv:2401.17263}, 2024{\natexlab{a}}.

\bibitem[Zhou et~al.(2024{\natexlab{b}})Zhou, Zhu, Yu, Li, Peng, Liu, and Han]{zhou2024modelinversionattackssurvey}
Zhanke Zhou, Jianing Zhu, Fengfei Yu, Xuan Li, Xiong Peng, Tongliang Liu, and Bo~Han.
\newblock Model inversion attacks: A survey of approaches and countermeasures, 2024{\natexlab{b}}.
\newblock URL \url{https://arxiv.org/abs/2411.10023}.

\bibitem[Zou et~al.(2023)Zou, Wang, Carlini, Nasr, Kolter, and Fredrikson]{zou2023universaltransferableadversarialattacks}
Andy Zou, Zifan Wang, Nicholas Carlini, Milad Nasr, J.~Zico Kolter, and Matt Fredrikson.
\newblock Universal and transferable adversarial attacks on aligned language models, 2023.
\newblock URL \url{https://arxiv.org/abs/2307.15043}.

\end{thebibliography}

\bibliographystyle{colm2025_conference}

\appendix

\section{Dataset statistics}
\hspace*{-3cm}
\begin{table*}[!h]
\centering
\resizebox{1.0\textwidth}{!}{   
    \begin{tabular}{l|c|c|c}  
        \midrule
        Dataset & Domain & \# of test example & \# of train example  \\
        \midrule
        BIPIA \citep{yi2023benchmarking} & code, email, table & 200 & 15 \\
        \hline
        SQuAD \citep{DBLP:journals/corr/RajpurkarZLL16} & Wikipedia & 400 & 5  \\
        \hline
        CaseHold \citep{zhengguha2021} & Law & 400 & 5 \\
        \hline
        FinQA \citep{DBLP:journals/corr/abs-2109-00122} & Finance & 400 & 5  \\
        \hline
        SciQ \citep{SciQ} & Science & 400 & 5 \\
        \hline
        TriviaQA \citep{joshi-etal-2017-triviaqa} & Open-domain & 400  & 5 \\
        \hline
        AQuA \citep{behrendt2024aquacombiningexperts} & Math & 400 & 5 \\
        \hline
        PubMedQA \citep{DBLP:journals/corr/abs-1909-06146} & Medical & 400 & 5 \\
        \bottomrule
    \end{tabular}
} 
\caption{Data Statistics}
\label{tab:data}
\end{table*}

The datasets cover a wide range of domains including Law, Finance, Science and so on, in order to evaluate prompt injection methods and defense models thoroughly. For most of domains, we collect 400 examples as test set and 5 examples as training set in our experiment, except BIPIA. We collect subset of BIPIA where the domains of \textit{code} and \textit{email} have 50 test examples individually and domain of \textit{table} has 100 test examples. For each subdomain of BIPIA,  we sample 5 examples in the corresponding training set as prompt injection training examples. 

\section{Training summary}
\begin{table*}[!h]
\centering
\resizebox{1.0\textwidth}{!}{   
    \begin{tabular}{l|c|c|c|c|c|c|c|c}  
        \hline
        Method & BIPIA & SQuAD & CaseHold & FinQA & SciQA & TriviaQA & AQuA & PubMedQA \\
        \hline
        GCG  & 0.64 (0.012)  & 0.6 (0.0)  & 0.667 (0.01)  & 1.0 (0.0)  & 0.933 (0.01)  & 0.2 (0.027)  & 1.0 (0.0) & 1.0 (0.0) \\
        AutoDAN & 0.38 (0.117) &  0.8 (0.0) & 0.33 (0.34) & 0.067 (0.094) & 0.6 (0.163)  & 0.4 (0.163)  & 0.53 (0.094)  &  0.533 (0.249) \\
        GBDA  & 0.576 (0.138) & 0.87 (0.163) & 0.4 (0.326) & 0.0 (0.0)  & 0.867(0.163)  & 0.4 (0.283) & 0.93 (0.094)  & 0.8 (0.163) \\
        AutoPrompt & 0.47 (0.18) & 0.6 (0.189) & 0.8 (0.0) & 0.53 (0.238) & 0.93 (0.094)  & 0.068 (0.094)  & 0.087 (0.094) & 0.087 (0.189)  \\
        PEZ  & 0.47 (0.158) & 0.667 (0.236) & 0.533 (0.236) & 0.4 (0.282) & 1.0 (0.0)  & 0.068 (0.094)  & 1.0 (0.0)  & 0.6 (0.282) \\
        UAT  & 0.71 (0.164) & 0.87 (0.189) & 0.867 (0.189) & 1.0 (0.0) & 0.867 (0.189)  & 0.2 (0.163)  & 0.93 (0.074)  & 0.93 (0.074) \\
        PAIR & 0.44 (0.182) & 0.93 (0.115) & 0.33 (0.231) & 0.6 (0.346) & 0.6 (0.2)  & 0.0 (0.0)  & 0.6 (0.2)  & 0.4 (0.2) \\
        \hline
    \end{tabular}
} 
\caption{Training ASR result}
\label{tab:method_comparison}
\end{table*}

We trained adverserial strings for each training data sample 3 times and calculate ASR of attacking training data sample with adverserial string. Then we average ASR and calculate standard deviation. 
\end{document}